\documentclass[prd,aps,eqsecnum,tightenlines,nofootinbib]{revtex4}
\input epsf
\usepackage{graphicx}

\newcommand{\beq}{\begin{equation}}
\newcommand{\beqa}{\begin{eqnarray}}
\newcommand{\eeq}{\end{equation}}
\newcommand{\eeqa}{\end{eqnarray}}

\newcommand{\lsim}{\lesssim}
\newcommand{\gsim}{\gtrsim}

\newcommand{\vect}[1]{\mbox{\boldmath${#1}$}}
\newcommand{\lmk}{\left(}
\newcommand{\rmk}{\right)}

\newcommand{\lkk}{\left[}
\newcommand{\rkk}{\right]}
\newcommand{\lla}{\left\langle}
\newcommand{\p}{\partial}
\newcommand{\rra}{\right\rangle}

\newcommand{\vex}{{\vect x}}

\newcommand{\ven}{{\vect \Omega}}

\newcommand{\vn}{{\vect n}}

\begin{document}
\title{Correlation analysis of  stochastic
gravitational wave background around 0.1-1Hz}
\author{Naoki Seto
}
\affiliation{Department of Physics and Astronomy, 4186 Frederick Reines
Hall, University of California, Irvine, CA 92697\\
Theoretical Astrophysics, MC 130-33, California Institute of Technology, Pasadena,
CA 91125
}

\begin{abstract}
We discuss  prospects for  direct measurement of  stochastic
 gravitational wave background around 0.1-1Hz with future space missions. 
It is assumed to use  correlation analysis technique with  the
 optimal TDI 
 variables for  two sets 
 of LISA-type interferometers.
The signal to noise for  detection of the  background
 and the  estimation  errors for its basic parameters
 (amplitude, spectral index)  are 
 evaluated for proposed  
missions.
\end{abstract}
\maketitle


\section{introduction}
Gravitational wave background generated in the early universe
is one of the most fascinating targets in observational cosmology \cite{Allen:1996vm,Maggiore:1999vm}.
Among others, inflationary theory has a realistic  mechanism to 
generate the background, and indeed 
confirmation of the background is regarded as
another strong support for presence of  inflationary phase in the early
universe. 
While we might indirectly detect the inflationary generated background
by B-mode polarization analysis of CMB \cite{Seljak:1996gy,Kamionkowski:1996zd}, direct detection of the
background with
gravitational wave detectors is  an indispensable approach to study the
inflation in more detail.   This is because the amplitude of the background
 is mainly determined
by the value of the inflation potential when the gravitational waves
cross the Hubble horizon in inflationary  epoch. If we can measure the
amplitudes at two widely  separated  frequencies ({\it e.g.} $\sim
10^{-17}$Hz and $\sim 0.1$Hz),  the global structure of the
potential might be constrained \cite{Lidsey:1995np, Turner:1996ck,bbo} (see also \cite{Seto:2003kc}).  The slope
of the 
spectrum at a given band will also provide us information of the
derivative of the potential. Therefore, it is quite meaningful to
understand how well we can measure the basic parameters that
characterize the spectrum.

Standard slow-roll inflation predicts that the spectrum $\Omega_{GW}(f)$
is nearly flat at frequency  regime  relevant for the direct
detection (\cite{Lidsey:1995np, Turner:1996ck}, see \cite{Ungarelli:2005qb,Cooray:2005xr,Smith:2005mm,Boyle:2005ug} for recent studies). This means that its strain amplitude is expected to be higher
at lower frequencies.  However,  astrophysical foreground would be  a
fundamental 
obstacle to directly detect  weak inflationary background below $\sim
0.1$Hz. For this reason the band around $\sim1$Hz is considered to be 
 suitable for the direct detection, and projects such as, the
Big Bang Observer (BBO; US) \cite{bbo} or DECI-hertz interferometer
Gravitational 
wave Observatory (DECIGO; Japan) \cite{Seto:2001qf}  have been proposed
(see also \cite{Ungarelli:2000jp,Crowder:2005nr}).
  For these projects,  correlation analysis   is a powerful
method to 
observe  weak background \cite{Michelson(1987),Christensen:1992wi,Flanagan:1993ix,Allen:1997ad}.
In this paper   prospects of this method are studied
 quantitatively, from
detectability of the background to  
parameter estimation errors.

This paper is organized as follows; In section II we study  basic
aspects of the optimal data streams for the Time-Delay-Interferometry
(TDI) 
method, and  designed sensitivities of BBO or DECIGO are briefly
mentioned. In section III a formal discussion for  correlation
analysis is presented. We evaluate the expected signal to  noise ratio
for detecting the background and derive  expressions for
parameter estimation errors based on the Fisher matrix approach.
In section IV numerical results for  BBO project are given with using
formulas in section III. Section V is a brief summary of this  paper.

\section{TDI variables and their responses to gravitational waves}
First we summarize  standard notations to discuss stochastic gravitational
wave background \cite{Allen:1997ad}.
The plane wave expansion for  gravitational waves  is given by 
\beq
h_{ab}(t,\vex)=\sum_{P=+,\times} \int^{\infty}_{-\infty} df \int_{S^2} d\ven
h_A(f,\ven) e^{2\pi i f (t-\ven \cdot \vex) } e^P_{ab}(\ven),
\eeq
where $f$ is the frequency of each  mode, $\ven$ is the unit vector for
its 
propagation, $S^2$ is a unit sphere for the angular integral $d\ven$, and $e^P_{ab}$ $(P=+,\times)$ is the basis for the
polarization tensor.  We assume that the stochastic background is
isotropic, unpolarized and static, and express the spectrum of its
amplitude 
$h_A(f,\ven)$ in terms of the logarithmic energy density of
the gravitational waves $\Omega_{GW}(f)\equiv 1/\rho_c
d\rho_{GW}(f)/d\ln f$ ($\rho_c$: the critical density) as follows; 
\beq
 \lla h_P^*(f,\ven) h_{P'}(f',\ven') \rra=\frac{3H_0^2}{32\pi^3}
\delta^2 (\ven-\ven')\delta_{PP'} \delta (f-f')|f|^{-3} \Omega_{GW}(|f|),\label{omega}
\eeq
where $H_0$ is the Hubble parameter and we fix it at $H_0=$70km/sec/Mpc.
The symbol $\lla ...\rra$  represents to take an ensemble average of 
 stochastic  quantities.

Next we discuss  responses of interferometers to incident
gravitational waves. We concentrate on a LISA-like detector.
Each unit is formed by three  spacecrafts at the vertices of a nearly
regular triangle, as shown in the solid lines in figure 1 where we also
define the labels of its vertices (1,2,3) and arms ($L_1,L_2,L_3$). 
We denote the six (one-way) relative frequency fluctuations of the laser
light as $y_{ij}(t)$ $(i,j=1,2,3; i\ne j)$.  The quantity $y_{ij}(t)$
corresponds to the signal measured at the spacecraft $j$, transmitted
from the spacecraft $k(\ne i,\ne j)$ along the arm $L_i$.
For example, the variable $y_{13}$ responds to a single  gravitational
wave mode 
with parameters $(f,\ven$ and $e^P_{ab})$ as \cite{estabrook}
\beq
y_{31}(t)=\frac12 \frac{ \vn_{12}\cdot  \vect{e}^P\cdot \vn_{12}}{1-(\ven\cdot\vn_{
12})^2}(1+\ven \vn_{12}) h_A(f,\ven) (U(t,1)-U(t-\tau,2)),
\eeq
where $\vn_{12}$ is the unit vector from vertex $1$ to vertex $2$, and  
the function $U(t,i)$ contains  information of the phase of the wave
at time $t$ and position $\vex_i$ of  vertex $i$.
 It is given as 
\beq
U(t,j)=\exp \lkk -2\pi i f (t-\ven\cdot \vex_j) \rkk.
\eeq
The time delay interferometry (TDI) is an important technique for
LISA-type
 detectors to overcome the laser frequency fluctuations \cite{Armstrong et al.(1999)}. We
follow Ref.\cite{Prince:2002hp} to summarize the relevant data streams
for signal 
analysis (see also \cite{Krolak:2004xp, Nayak:2002ir,Corbin:2005ny}).  We first define a TDI variable, $\alpha$ as follows;
\beq
\alpha=y_{21}-y_{31}+y_{13,2}-y_{12,3}+y_{32,12}-y_{23,13},
\eeq
where we used the notations like $y_{13,2}(t)\equiv y_{13}(t-L_2)$ and
$y_{32,12}(t)\equiv y_{32}(t-L_1-L_2)$.
In the same  manner we define two other TDI variables $\beta$ and $\gamma$
that are given by  cyclic permutations of the subscripts of the variable $\alpha$.
These $\alpha$, $\beta$ and $\gamma$ are TDI variables, but their noises
are correlated. Thus we define the following new variables $A$, $E$ and
$T$ as
\beqa
A&=&\frac1{\sqrt 2}(\alpha-\gamma),\\
E&=&\frac1{\sqrt 6}(\alpha-2\beta+\gamma),\\
T&=&\frac1{\sqrt 3}(\alpha+\beta+\gamma).
\eeqa
As we can easily confirm with using the symmetry of the original data
streams $\alpha$, $\beta$ and $\gamma$, the noises of the variables $A$, $E$
and $T$ do not have correlation.  In other words, they are
orthogonal. We regard them as 
the fundamental data  sets for correlation analysis. Hereafter, we do
not discuss  differences of  the arm lengths 
$L_i$ or their time
variations, and simply put
$L_1=L_2=L_3=L=const$.  Even if  the second generation TDI
variables \cite{Cornish:2003tz,Shaddock:2003dj} are used, our
basic results are not changed \cite{Krolak:2004xp}. 
As our primary interest is   
observation of the monopole mode of the background,  the motion of the triangle will not be  included.

\begin{figure}
\begin{center}
\includegraphics[scale=0.4]{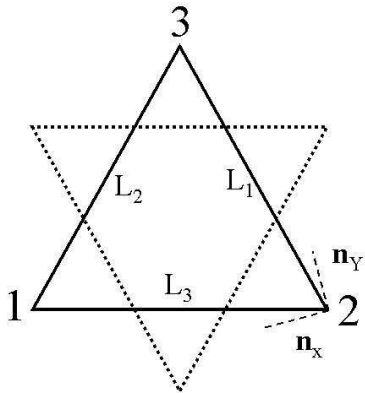} 
\end{center}
\caption{Three vertices $(1,2,3)$ and three arms $(L_1,L_2,L_3)$ of
 the first unit (solid lines). Configuration of  second unit
 (short-dashed  lines) is obtained  by $180^\circ$ rotation of the first
 one around its center. Labels for vertices and arms of the second ones
are transported with this rotation.}
\label{f1}
\end{figure}

At low frequency regime with $fL\ll 1$, the response of $E$ mode is
approximately given as \cite{Krolak:2004xp,white}
\beq
E\simeq \frac3{\sqrt 2} (2\pi f L)^2 \lkk\frac12(\vn_X \cdot {\vect
h}\cdot \vn_X-\vn_Y\cdot {\vect h}\cdot \vn_Y)   \rkk, \label{lf} 
\eeq
where directions of two unit vectors $\vn_X$ and $\vn_Y$ ($\vn_X\perp\vn_Y$) are
given in figure 1. The large parenthesis in eq.(\ref{lf}) is the response
of a simple $L$-shaped detector, as often used in the literature of
gravitational waves.  In the same manner, the $A$ mode 
asymptotically becomes a simple response obtained by rotating
$\vn_X-\vn_Y$ system by $45^\circ$ on the detector plane
\cite{Cutler:1997ta}.   The  response 
of   T
mode at low frequency regime is $T=O(f^3)$ in contrast to $A=E=O(f^2)$
as in eq.(\ref{lf}).  
This allows us to use  the T mode to monitor the detector noise in
principle \cite{Tinto:2001ii,Hogan:2001jn}.

As for the sources of noises,  we take into account the proof mass and optical
path noises with parameters $w_p$ and $w_o$ as
\beqa
S_y^{proof-mass}(f)&=&2.5\times 10^{-48}w_p^2 (f/1{\rm Hz})^{-2}  {\rm Hz^{-1}},\label{proof}\\
S_y^{optical-path}(f)&=&1.8\times 10^{-37}w_o^2 (f/1{\rm Hz})^{2}  {\rm Hz^{-1}}.\label{optical}
\eeqa
The proof mass noise is dominant at low frequency regime.
Basic parameters of LISA are $(L,w_p,w_o)=(5.0\times 10^6{\rm km},
1.0,1.0)$ \cite{lisa,Prince:2002hp,white}. For  BBO project 
three possible configurations are discussed \cite{bbo},  namely BBO-lite: $(2.0\times
10^4{\rm km}, 0.1, 4.0\times10^{-6})$, BBO-standard: $(5.0\times
10^4{\rm km}, 0.01, 1.4\times10^{-6})$ and BBO-grand 
 $(2.0\times
10^4{\rm km}, 0.001, 2.6\times10^{-7})$ \cite{bbo}.
These parameters for BBO configurations were presented in 2003, and we use
them for a reference. We should keep in mind that they do not always
 represent the most up to date designed sensitivities.  In the original proposal the
second configuration; 
``BBO-standard'' is simply named as ``BBO''. In this paper 
 we use the former for the name of 
a specific configuration and the latter for the name of  the
project itself.    
 The parameter $w_0$ is assumed
to scale as
$w_0\propto \lambda^{3/2} (\epsilon p)^{-1/2}LD^{-2} $ with the laser
power $p$ at wavelength $\lambda$, optical efficiency $\epsilon$, the
armlength $L$ and the mirror diameter $D$ \cite{Larson:1999we}.
The noises (\ref{proof}) and (\ref{optical}) are defined for the one-way 
signal $y_{ij}$. For the A, E and T modes  the noise spectra are given
by \cite{Prince:2002hp,Krolak:2004xp,white}
\beqa 
S_A(f)&=&S_E(f)=16\sin^2(f/2f_*) \lkk 3+2\cos(f/f_*) +\cos(2f/f_*) \rkk
S_y^{proof-mass}(f)\nonumber\\
& &+8\sin^2(f/2f_*) \lkk
2+\cos(f/f_*)\rkk S_y^{optical-path}(f)  ,\label{ae}\\
S_T(f)&=&2 \lkk 1+2\cos(f/f_*)  \rkk^2
\lkk 4\sin^2(f/2f_*)S_y^{proof-mass}(f)+S_y^{optical-path}(f)
\rkk\label{t} 
\eeqa
with $f_*\equiv (2\pi L)^{-1}$. We have $f_*=0.95$Hz for
BBO-standard and 
$f_*=2.4$Hz for BBO-grand and BBO-lite. 

As the responses $H_I$  ($I=A,E,T$)  to the
coefficient  $h(f,\ven)$ of each gravitational wave mode are  linear, we can express
them in a form
\beq
H_I=h_P(f,\ven) R(I, f,\ven,P).\label{hi}
\eeq
After taking the average with respect to the direction $\ven$ and polarization $P$
of the incident waves, we obtain the effective noise curve
\cite{Prince:2002hp,Larson:1999we,Cornish:2001qi} 
\beq
h_I(f)=\lmk \frac{\lla R^*(I,f,\ven,P) R(I,f,\ven,P) \rra_a}{S_I(f)}
\rmk^{-1/2}, \label{noise}
\eeq
where the symbol $\lla ... \rra_a$ represents the above mentioned average.
Due to  symmetry of the data streams ($A,E,T$) we have $\lla
R^*(I,f,\ven,P) 
R(I,f,\ven,P) \rra_a=0$ for $I\ne J$.  This means that the correlation
between ($A,E$), ($E,T$) and ($T,A$) vanish for isotropic
component of the background.
Thus the optimal sensitivity to the isotropic gravitational wave background with  all
the three variables becomes
\beq
h_{opt}(f)=\lmk \sum_{I=A,E,T}\frac{\lla R^*(I,f,\ven,P) R(I,f,\ven,P) \rra_a}{S_I(f)}
\rmk^{-1/2}.
\eeq 
In figure 2 we show some of the results for the  three possible BBO
configurations (see also \cite{Crowder:2005nr}). The $A$ and $E$ modes have
the same  sensitivity. 
In the low frequency regime $f\ll f_*$ the contribution of the T-mode is
negligible to  the optimal sensitivity, and we have $h_{opt}(f)\simeq
h_A(f)/\sqrt{2}$ \cite{Prince:2002hp}. The currently designed
sensitivity of    Fabry-Perot 
type DECIGO is similar 
to that of BBO-standard \cite{kawamura}. While the minimum noise floor of DECIGO extends
to a higher frequency ($\sim$7.5Hz) than  BBO-standard (see figure 2),
this 
difference is not important for observing the inflationary background
and we will have similar results for these two cases (in order of
magnitude sense).
Therefore we do not take up   DECIGO separately from BBO-standard.

\begin{figure}
\begin{center}
\includegraphics[scale=0.4]{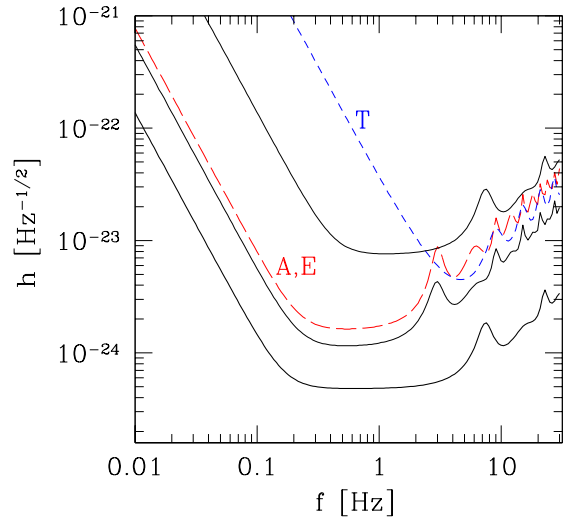} 
\end{center}
\caption{The optimal sensitivities for BBO-lite (thin solid curve),
 BBO-standard 
 (solid curve) and 
 BBO-grand (thick solid curve) configurations. The sensitivities for (A,E) modes
 (long-dashed curve) and T mode (short-dashed curve) are also given for
 BBO-standard configuration.} 
\label{f2}
\end{figure}

\section{correlation analysis with star-like configurations}
Correlation analysis is a powerful approach to detect weak stochastic
gravitational wave
background \cite{Michelson(1987),Christensen:1992wi,Flanagan:1993ix,Allen:1997ad}. If noises of two data streams have no  correlation
but their responses to gravitational wave background are correlated,
we can increase the signal to noise ratio for detection of background by
a long term 
observation. 
In the case of LISA  the orthogonal data
streams $A$, $E$ and $T$ do not have correlated responses as mentioned in
the last section, and we cannot perform   correlation analysis to
measure the monopole mode of the background. 
The situation is different  for studying  anisotropies of the
background, and the correlation between two of the three ($A,E,T$) modes
will  
be useful to extract information of the background generated 
by  Galactic
binaries \cite{Seto:2004ji,Kudoh:2004he,Seto:2004np}.  
For example, we can show that correlation between $A$ and $E$ modes has
sensitivity to the hexadecapole  mode ($l=4$)  under the low frequency
approximation. By combining this effect with the time modulation of the
signal due 
to the rotation of the detector, we can get information of low-$l$
moments ({\it e.g.} $l=2,4$) of the Galactic gravitational wave
background around $\sim1$mHz that is expected to be highly anisotropic 
\cite{Seto:2004ji,Kudoh:2004he,Seto:2004np}.

For BBO  it is proposed to use another unit for   correlation
analysis to detect isotropic mode.  In figure 1 we show the proposed placement of its two
units \cite{bbo,Crowder:2005nr,Cornish:2001qi}. Two units have identical
specification. The  
position of  the second unit (dotted lines) is obtained by rotating the
first one by $180^\circ$ degree with respect to  the center of the triangle. 
We  transport the labels of the first one to the second one
$(1',2',3',L_1',L_2',L_3')$ 
 with this rotation. For example, the vertex 2' is at the opposite side
of the vertex 2 around the center. The orthogonal TDI variables $A', E'$
and $T'$ of the second unit are defined in the same  manner as the first
one in the last section.

To discuss the correlated response of two variables $I$ and $J$,  we
define the overlap reduction function $\gamma_{IJ}(f)$ as \cite{Flanagan:1993ix,Allen:1997ad,Cornish:2001qi}
\beq
 \gamma_{IJ}(f) \equiv 5 \lla R^*_I R_J \rra_a.\label{red}
\eeq
Due to  symmetry of the relevant data streams, we have $ 5 \lla R^*_I
R_J \rra_a =5 \lla R_I R^*_J \rra_a$ and the functions  $\gamma_{IJ}$
take  real numbers.
Furthermore, their  nonvanishing combinations 
 $\gamma_{IJ}$ $(I\ne J)$ to isotropic mode are only  $\gamma_{AA'}=\gamma_{EE'}$ and
$\gamma_{TT'}$. Other combinations ({\it e.g.} $\gamma_{AE'}$,
$\gamma_{TA'}$) become zero.  As in the case of LISA, we can, in
principle, study anisotropic components ($l\ge 1$) by correlating two
outputs, such as, $(A,E)$ or $(A,E')$. However, it would be difficult to
observe the cosmological anisotropy of the background around $\sim 1$Hz,
considering its 
expected magnitude $\sim 10^{-5}$ \cite{Seto:2004np} (see also
\cite{Kudoh:2005as}).  
The factor 5 in eq.(\ref{red}) is the conventional choice \cite{Flanagan:1993ix,Allen:1997ad,Cornish:2001qi}, and  we have
$\gamma_{IJ}=1$ for two co-aligned detectors at low frequency limit with
the 
simple response function in the large parenthesis of eq.(\ref{lf}).  
In figure 3 we show these three non-vanishing overlap reduction functions
normalized by 
$9(f/f_*)^4/2$ that comes for the prefactor in eq.(\ref{lf}).
These functions are determined purely by the geometry of the
placement and scale with the parameter $f_*=1/2\pi L$ for any
star-like constellation given  in figure 1.

\begin{figure}
\begin{center}
\includegraphics[scale=0.4]{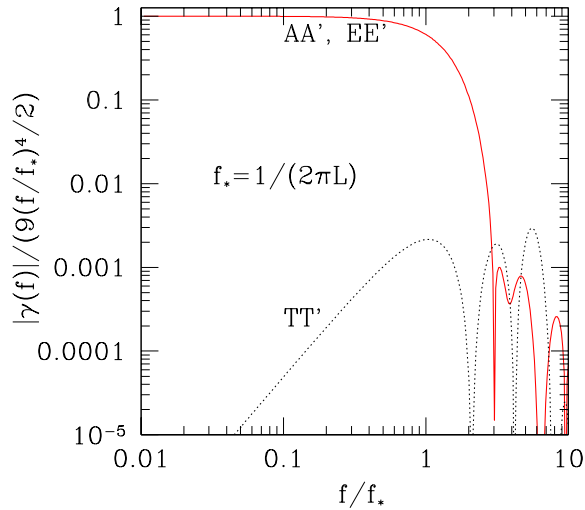} 
\end{center}
\caption{The normalized overlap reduction function $\gamma(f)$ for a star-like constellation. These
 curves have a scaling parameter $f_*=c/(2\pi L) $ that is determined by the arm-length.}
\label{f3}
\end{figure}

Now we study correlation analysis in a more detailed manner with Fourier
space 
representation. Each data
stream $s_I(f)$ is made by gravitational wave signal $H_I(f)$ and 
noise $n_I(f)$ as
\beq
s_I(f)=H_I(f)+n_I(f).
\eeq
The noise spectrum 
\beq
\lla n_I^*(f) n_I(f')\rra=\frac12\delta(f-f')S_I(f)
\eeq
is given in eqs.(\ref{ae}) and (\ref{t}) for the orthogonal TDI
variables $I=\{A,E,T,A',E',T'\}$. We assume that the noises have 
no correlation (namely $\lla n_I^* n_J\rra=0$ for $I\ne J$), and the amplitude of the signal $\lla H_I(f)^* H_J(f) \rra$
is much smaller than that of  the noise   $\lla n_I(f)^* n_I(f)
\rra$. The latter 
is the condition where  correlation analysis becomes very  powerful.
We divide the positive  Fourier space into frequency segments $F_i$
($i=1,...,N$) 
with their center frequencies $\{f_i\}$ and widths $\{\delta f_i\}$. 
 In each segment the width $\delta f_i$ is much smaller
than $f_i$, and the relevant quantities (e.g. $\Omega_{GW}(f)$,
$\gamma_{IJ}(f)$) are almost constant. But the width is
much larger than the frequency resolution $\Delta f\equiv T_{obs}^{-1}$ 
($T_{obs}$: observation period) so that each  segment contains 
Fourier modes as many as $\delta f_i/\Delta f\gg 1$. 

For 
correlation analysis we compress the
observational data $s_I(f)$ by summing up the products $s^*_I(f)s_J(f)$ 
 ($I\ne J$) in each segment $F_i$ as
\beq
\mu_i\equiv\sum_{f\in F_i} s^*_I(f)s_J(f),\label{mean}
\eeq
where we omitted the apparent subscript $\{IJ\}$ for the compressed
data  $\{\mu_i\}$ for
notational 
simplicity. As the noises are assumed to be uncorrelated, the statistical
mean 
$\lla \mu_i \rra$ is arisen by  gravitational wave
signal. After some calculations with using eqs.(\ref{omega}) (\ref{hi})
and (\ref{red}) we have a real value
\beq
\lla \mu_i\rra= \sum_{f\in F_i}\lla  H_I(f)^*H_J(f)\rra\simeq \frac{3H_0^2}{20\pi^2} f_i^{-3} \Omega_{GW}(f_i) \gamma_{IJ}(f_i)\frac{\delta
f_i}{\Delta f}. \label{variance}
\eeq 
The fluctuations around the mean $\lla \mu_i\rra$ are dominated by the
noise under our weak signal approximation,  and its variance
$\sigma_i^2$ for  the real part of $\mu_i$  becomes 
\beq
\sigma_i^2=S_I(f_i) S_J(f_i) \frac{\delta f_i}{8\Delta f}.
\eeq
As the number of Fourier modes $\delta f_i/\Delta f$ in each segment
is much larger 
than unity, the probability distribution function (PDF) for the real part of  the measured
value $\mu_i$ is  close to 
Gaussian distribution due to the central limit theorem as
\beq
p({\rm Re} [\mu_i])\simeq\frac1{\sqrt{2\pi \sigma_i^2}}\exp \lkk-\frac{({\rm Re} [\mu_i]-\lla \mu_i\rra)^2}{2\sigma_i^2} \rkk. \label{pdf}
\eeq
Here we neglected the prior
information of the spectrum $\Omega_{GW}(f)$.
From eqs.(\ref{mean}) and (\ref{variance}) the signal to noise ratio of each segment becomes
\beq
SNR^2_i=\frac{\lla \mu_i\rra^2 }{\sigma_i^2}=\lmk
\frac{3H_0^2}{10\pi^2}\rmk^2  {T_{obs}} \lkk 2\delta f_i
\frac{\gamma_{IJ}(f)^2 \Omega_{GW}(f)^2}{f^6 S_I(f)S_J(f)}   \rkk. 
\eeq
Summing up the all the segments quadratically, we get the total signal to noise ratio 
\beq
SNR^2= \lmk \frac{3H_0^2}{10\pi^2}\rmk^2 T_{obs} \lkk2 \int_0^\infty df
\frac{\gamma_{IJ}(f)^2 \Omega_{GW}(f)^2}{f^6 S_I(f)S_J(f)}   \rkk. \label{single}
\eeq
Note that this expression does not depend on the details of the
segmentation  
$\{F_i\}$. We can directly obtain the same  results  by  introducing the
optimal filter for 
the product $s_I^*(f) s_J(f)$ to get the highest signal to  noise ratio
(see {\it e.g.} \cite{Allen:1997ad}).
Eq.(\ref{single}) is given for a single pair $(I,J)$ of the data
streams. As the overlap reduction functions $\gamma_{IJ}$ are 
``diagonalized'' ($\gamma_{IJ}\ne0$ only for the following combination
$(I,J)=(A,A'),(E,E'), 
(T,T')$ with $I\ne J$), the total signal to noise ratio given by all of
these 
combinations is evaluated  by  adding their  contributions as
\beq
(SNR)^2_{opt}=\lmk \frac{3H_0^2}{10\pi^2}\rmk^2 T_{obs} \lkk2 \sum_{(I,J)}\int_0^\infty df
\frac{\gamma_{IJ}(f)^2 \Omega_{GW}(f)^2}{f^6 S_I(f)S_J(f)}
\rkk. \label{optimal} 
\eeq 
In this paper we mainly study the case with using all these three
combinations, unless otherwise stated.

Next we discuss how well we can estimate  parameters $\alpha_m$
($m=1,...,M$; $M$: total number of parameters) that characterize
the stochastic background spectrum $\Omega_{GW}(f)$. 
We can apply  standard procedure of the maximum likelihood analysis
for  the compressed data $\{\mu_i\}$ with the probability distribution
function (\ref{pdf}) \cite{helstrom}.  Then the
magnitude of the  
 parameter estimation error $\Delta \alpha_m$
is evaluated by the Fisher information matrix $\Gamma_{mn}$ that is the
inverse of the error covariance matrix $\lla \Delta \alpha_m \Delta
\alpha_n\rra$  as
\beq
\lla \Delta \alpha_m \Delta\alpha_n\rra^{-1}=\Gamma_{mn}
=\lmk \frac{3H_0^2}{10\pi^2}\rmk^2 T_{obs} \lkk 2
\sum_{(I,J)}\int_0^\infty df 
\frac{\gamma_{IJ}(f)^2 \p_{\alpha_m}\Omega_{GW}(f)
\p_{\alpha_n}\Omega_{GW}(f)  }{f^6 S_I(f)S_J(f)}
\rkk. \label{error} 
\eeq

The simplest case is the estimation of the amplitude $\Omega_{GW}$ for
a flat spectrum $\Omega_{GW}(f)=\Omega_{GW}=const$ in the 
frequency range  relevant for  correlation analysis. In this one
parameter 
estimation with $\alpha_1=\ln \Omega_{GW}$,  the  expected error  becomes
\beq
\Delta \alpha_1=\frac{\Delta \Omega_{GW}}{\Omega_{GW}}=(SNR)^{-1}.\label{one}
\eeq
 We can easily confirm  this by  using
eqs.(\ref{optimal}) and (\ref{error}).
Actually,  the above result 
(\ref{one}) holds for the case when we  estimate only the
overall amplitude of the spectrum $\Omega_{GW}(f)$ with a known
frequency dependence. 

A more realistic situation is to  estimate two parameters, the amplitude
$\alpha_1=\ln \Omega_{GW,F}$ and the slope $\alpha_2=n$, assuming that the spectrum
has a power-law form 
\beq
\Omega_{GW}(f)=\Omega_{GW,F}(f/F)^n
\eeq
 around a
central frequency $F$. In this case we do not have the simple result
(\ref{one}) for the first  parameter  $\alpha_1=\ln \Omega_{GW,F}$, as the two
parameters have 
correlation. The magnitude of the error $\Delta \alpha_2$ for the slope
does not depend 
on the choice 
of the frequency $F$, as we can understand from its geometrical
meaning. We can also confirm this directly with using eq.(\ref{error}).
In  contrast, the error $\alpha_1$ and the correlation coefficient
$r\equiv {\lla \Delta \alpha_1\alpha_2 \rra}/({\lla \Delta
\alpha_1^2 \rra \lla \Delta \alpha_2^2 \rra})^{1/2}$ depend on the
frequency 
$F$. With a suitable choice of $F$ we can diagonalized the covariance
matrix $\lla \Delta \alpha_i\alpha_j \rra$ $(i=1,2)$.
Once we get the errors $\Delta \alpha_i$ or signal to
noise ratio for a specific combination $(\Omega_{GW,F},n)$, we can
easily obtain 
the results for a different amplitude $\Omega_{GW,F}$ (but the same slope
$n$)  with using  scaling relations as
\beq
SNR\propto \Omega_{GW,F},~~~\Delta \alpha_i \propto
\Omega_{GW,F}^{-1},~~~ r\propto \Omega_{GW,F}^0. \label{scale}
\eeq
It is straightforward to confirm these relations with
eqs.(\ref{optimal}) and (\ref{error}).

So far we have assumed the situation (weak-signal approximation) that
the detector noise $\{n_I(f)\}$ is much larger than  the gravitational wave
background 
 $\{H_I(f)\}$. When the latter becomes comparable to the former, the
background $H_I(f)$ itself becomes a  significant source of the fluctuations for
measuring $\lla \mu_i\rra$ in eq.(\ref{variance})
\cite{Allen:1997ad,Kudoh:2005as}.  For the strong 
signal limit, the signal to noise ratio becomes a
asymptotic value of order $SNR\sim (T_{obs} f)^{1/2}$ that is $10^3\sim
10^4$ for BBO band with a reasonable observational time. In a recent paper
\cite{Kudoh:2005as} it is shown that for BBO band ($\gsim 0.1$Hz) the
weak signal approximation provides a fairly good estimation for SNR
smaller than $\sim 200$.  On the other hand, the performance of the
Fisher matrix approach becomes worse at low SNR. 
Therefore,  our expressions 
in this section are expected to be reasonable  for  the background observed
with 
$10\lsim SNR\lsim 200$.

\section{Numerical results for BBO}
In  this section we present numerical results for  BBO project. Its
primary goal is  direct detection of the stochastic gravitational
wave background generated at inflation. As the standard slow-roll
inflation predicts a nearly flat spectrum at
BBO band,  we first assume that the true
spectrum has a simple form: $\Omega_{GW}(f)=\Omega_{GW}=const$. 
By evaluating eq.(\ref{optimal}) numerically, we obtain the signal to noise
ratio for three possible BBO configurations as 
\beqa
SNR&=&1.03 \lmk \frac{\Omega_{GW}}{10^{-15}} \rmk \lmk
\frac{T_{obs}}{10 {\rm yr}} \rmk^{1/2}~~~({\rm BBO-lite}),\label{lite}\\
SNR&=&25.1 \lmk \frac{\Omega_{GW}}{10^{-16}} \rmk \lmk
\frac{T_{obs}}{10 {\rm yr}} \rmk^{1/2}~~~({\rm
BBO-standard}),\label{standard}\\ 
SNR&=&251.2 \lmk \frac{\Omega_{GW}}{10^{-16}} \rmk \lmk
\frac{T_{obs}}{10 {\rm yr}} \rmk^{1/2}~~~({\rm BBO-grand})\label{grand}.
\eeqa 
The magnitude $\Omega_{GW}=10^{-15}$ is close to the upper limit of the
amplitude around the  BBO band that is consistent with the current CMB
observation \cite{Turner:1996ck}. While specification of  BBO-lite is not enough
to detect the 
level $\Omega_{GW}\sim 10^{-15}$ with a sufficient signal
to  noise ratio,  BBO-grand has potential to detect the background
close to $\Omega_{GW}\sim 10^{-18}$.  

To calculate  the above results we simply integrated  eq.(\ref{optimal}) from
$f=0$ to $f=\infty$. This would be too optimistic  considering the
fact that the frequency below $f\lsim 0.2$Hz might be significantly
contaminated by cosmological white dwarf binaries
\cite{Farmer:2003pa}. Above $\sim 0.2$Hz we 
still have to clean the foreground produced by the binaries made by
neutron stars or black holes \cite{Seto:2001qf,Ungarelli:2000jp}. While
it is not clear how well we can 
actually
perform this cleaning (see \cite{Cutler:2005qq} for recent study), we
calculate a less optimistic prediction than 
eqs.(\ref{lite})(\ref{standard}) and (\ref{grand}) by
introducing a lower frequency cut-off $f_{cut}$ at $f_{cut}=0.2$Hz for
the integral (\ref{optimal}).
Then the above relations  become
\beqa
SNR&=&0.98 \lmk \frac{\Omega_{GW}}{10^{-15}} \rmk \lmk
\frac{T_{obs}}{10 {\rm yr}} \rmk^{1/2}~~~({\rm BBO-lite}),\label{lite2}\\
SNR&=&17.3 \lmk \frac{\Omega_{GW}}{10^{-16}} \rmk \lmk
\frac{T_{obs}}{10 {\rm yr}} \rmk^{1/2}~~~({\rm
BBO-standard}),\label{standard2}\\ 
SNR&=&116.0 \lmk \frac{\Omega_{GW}}{10^{-16}} \rmk \lmk
\frac{T_{obs}}{10 {\rm yr}} \rmk^{1/2}~~~({\rm BBO-grand})\label{grand2}.
\eeqa 
In figure 4 we show how the signal to noise ratio changes with this
cut-off frequency. BBO-lite is less
sensitive to $f_{cut}$, compared with other two configurations. This is
because its minimum of the 
noise curve is at higher frequency than BBO-standard or BBO-grand, as in
figure 2.   The upper cut-off frequency is not important for our
results, if it is higher than $\sim 1$Hz. Therefore we do not discuss
its effects.
From  figure 2 we can expect that the contribution of
T-mode to the total SNR is small. Actually, even if we remove ($T,T'$) correlation from
eq.(\ref{optimal}),  the  prefactors in eqs.(\ref{lite2})(\ref{standard2}) and (\ref{grand}) change
less than 1\%.
We also study the case with completely aligned two units on a single
triangle for correlation analysis. The total SNR is obtained by putting
$\gamma_{IJ}=5\delta_{IJ}$ and $S_I(f)=h_I(f)^2$ in eq.(\ref{optimal}).
Here $h_I(f)$ is the effective noise curve for $I$-mode given in figure
2. We find that the above prefactors become 0.98, 17.4 and 116.1
respectively, and are very close to the results with the proper transfer
functions. Therefore, we can approximately discuss performance of a
star-like constellation in  a very convenient manner with the effective
noise curves that are often used to represent specification of an
interfereometer.

\begin{figure}
\begin{center}
\includegraphics[scale=0.4]{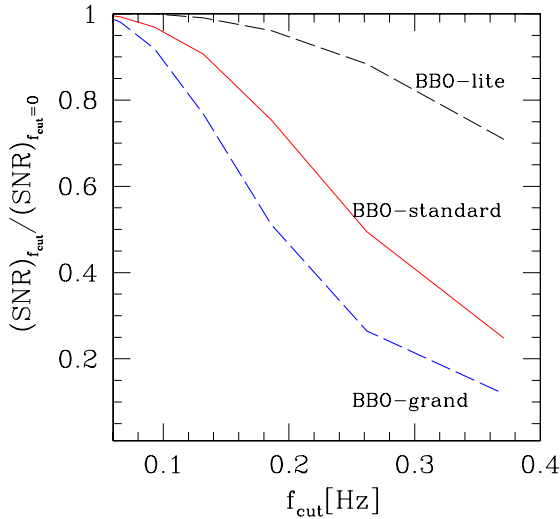} 
\end{center}
\caption{Dependence of the signal to noise ratio on the lower cut-off
 frequency $f_{cut}$.}
\label{f4}
\end{figure}

\begin{figure}
\begin{center}
\includegraphics[scale=0.6]{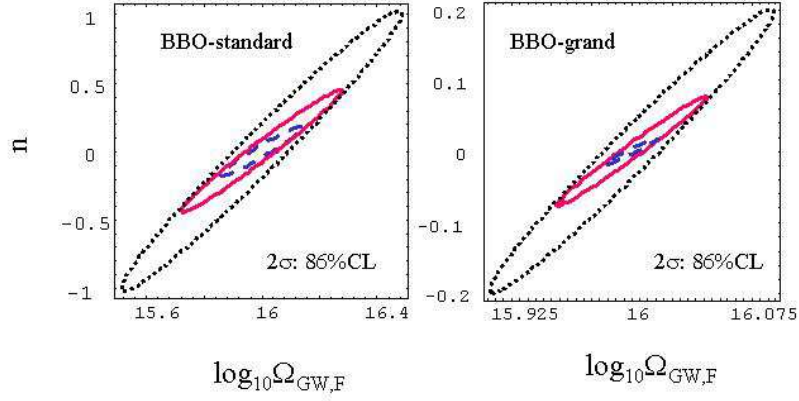} 
\end{center}
\caption{Error ellipses ($2\sigma$; 86\%CL) for two dimensional
 parameter estimation with different lower cut-off frequencies $f_{cut}$
 (dashed curve: $f_{cut}=0$,  solid curve: $f_{cut}=0.2$Hz, 
  the dotted curve: $f_{cut}=0.3$Hz). The fisher matrix approach is used.  The central frequency $F$ is set at
 $F=1$Hz that is slightly higher than the optimal sensitivity for
 measuring 
 the background with above two BBO configurations.  The  true values of the
 spectrum $\Omega_{GW}(f)$  are 
   $\Omega_{GW,F}=10^{-16}$ and $n=0$.  }
\label{f5}
\end{figure}

\begin{figure}
\begin{center}
\includegraphics[scale=0.6]{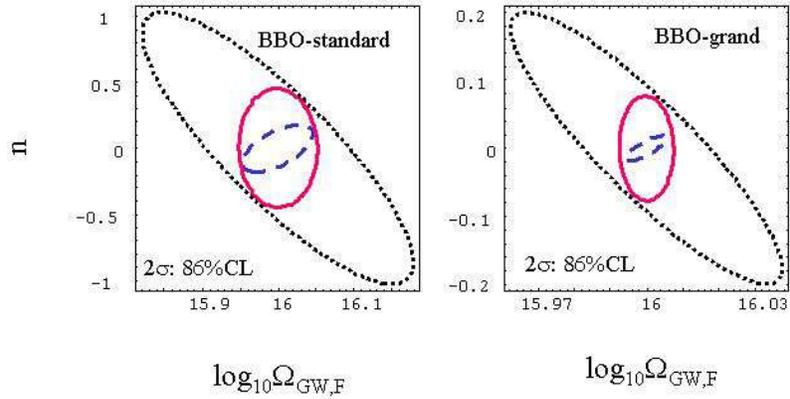} 
\end{center}
\caption{ Same as figure 5, but with the central  frequencies $F$ at
 $0.26$Hz (BBO-standard) and $0.25$Hz (BBO-grand). The two
 dimensional covariance matrices are diagonalized with $f_{cut}=0.2$Hz.}
\label{f6}
\end{figure}

Now we move to  the case with estimating two parameters $(\alpha_1,\alpha_2)=(\ln\Omega_{GW,F},n)$ for the assumed spectrum shape
$\Omega_{GW}(f)=\Omega_{GW,F} (f/F)^n$. 
We fix the true values of the parameters at $n=0$ and 
$\Omega_{GW,F}=10^{-16}$, and set the lower cut-off frequency $f_{cut}$
at 0.2Hz.  As we discussed in the last section, the estimation error for
the slope $n$ does not depend on the choice of the central frequency
$F$, and we get
\beq
\lla \Delta \alpha_2^2 \rra^{1/2} \simeq
\Delta n \simeq 0.23~~~(0.039),\label{e2}
\eeq
for BBO-standard (BBO-grand, respectively). In contrast, the error
$\Delta \alpha_1$ for the amplitude $\Omega_{GW,F}$ and the correlation
coefficient $r$ depend on the frequency $F$. From  observed data we
can determine the profile of the spectrum
$\Omega_{GW}(f)$  relatively well around  the optimal frequency  region
where the signal to noise ratio accumulates in
eq.(\ref{optimal}). However, 
if we take  the central frequency $F$ away from this optimal region, the
estimated amplitude $\Omega_{GW,F}$ at the frequency $F$ would be
strongly affected by 
the error of the slope $n$. Consequently, the correlation between errors
of the  two
parameters $(\ln \Omega_{GW,F},n)$ becomes strong.
In figure 5 we show $2\sigma$-contour map expected for the two parameter
fitting.
In this figure we take $F=1$Hz that is higher than the optimal frequency
region around $0.2\sim 0.4$Hz (see figure 4). We can observe a strong
correlation  between two fitting parameters, and we have $|r|>0.96$ for
the results given in figure 5.

We searched the central frequency $F$ that makes the correlation
coefficient $r=0$, and found $F=0.26$Hz (BBO-standard) and 0.25Hz
(BBO-grand) for the lower cut-off frequency at 0.2Hz.
With these choices for the frequency $F$, the error $\Delta \alpha_1$
for the amplitude  
are given by $SNR^{-1}$ as eq.(\ref{one}),  as the two dimensional
variance matrix $\lla 
\Delta\alpha_i\Delta\alpha_j\rra$ is now diagonalized.
In figure 6 we plot the error ellipses with these frequencies $F$. 
Compared with figure 5, the error for the amplitude $\Omega_{GW,F}$
becomes significantly smaller. We can  also understand the direction of
the ellipse (or the sign of the coefficient $r$) with an argument
similar to the discussion given just after eq.(\ref{e2}).

So far we have studied the results mainly for a specific model
$\Omega_{GW}(f)=10^{-16}$ at the BBO band.
Here we analyze how various quantities depend on the spectral
index $n$ for the two dimensional parameter fitting ($\alpha_1=\ln
\Omega_{GW,F}$, $\alpha_2=n$)  with assumed spectral form $\Omega_{GW}=
\Omega_{GW,F}(f/F)^n$. We fix $\Omega_{GW,F}=10^{-16}$, $f_{cut}=0.2$Hz,
$F=0.26$  for 
BBO-standard, 0.25Hz for BBO-grand, 
 but change  the slope $n$.
 Then the signal to noise ratio, the magnitude of errors $(\Delta
\alpha_1, \Delta \alpha_2)$,
and their 
correlation coefficient $r$ between them are evaluated as functions of
the slope $n$.
The results are presented in figure 7. Note that the signal to noise
ratio and the error for the amplitude $\Omega_{GW,F}$ depend very weakly
on the slope.  This is because the central frequency $F$ is  in
the optimal frequency  region. With this figure and the numerical results
given so far,  we can  get relevant
quantities  for various combinations $(\Omega_{GW,F},n)$
by using the scaling relations 
given in eq.(\ref{scale}).

\begin{figure}
\begin{center}
\includegraphics[scale=0.4]{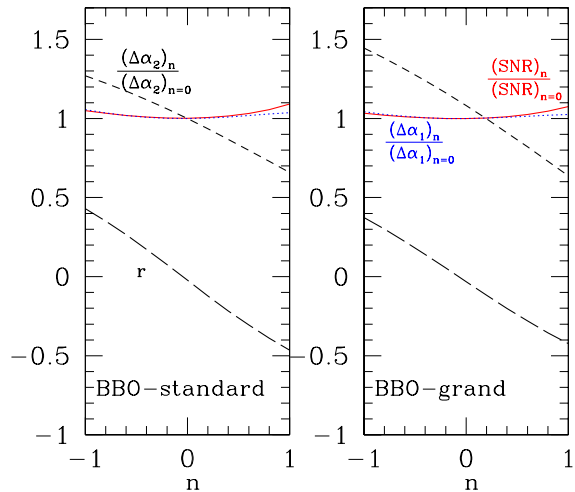} 
\end{center}
\caption{Dependence of various quantities to the slope $n$. SNR
 (solid curve) and two errors $\Delta \alpha_1$ (dotted curve) and
 $\Delta \alpha_2$ (short-dashed curve)  are
 normalized by their values at $n=0$.
 The long-dashed curve represent the correlation coefficient $r$.
The central frequencies   
are  $F=0.26$Hz (BBO-standard) and $F=0.25$Hz (BBO-grand).}
\label{f7}
\end{figure}

\section{summary}

In this paper we studied  prospects for the direct measurement of the
stochastic gravitational wave background by  correlation analysis. As
a concrete example, we explicitly examined the case with possible BBO
configurations that use two sets of three spacecrafts to form a
star-like constellation.  We calculated not only the
signal to noise ratio for detection, but also how accurately we can
measure the basic parameters that characterize the spectrum
$\Omega_{GW}(f)$, such as its amplitude $\Omega_{GW,F}$ or its slope $n$.

While it is difficult to detect a level $\Omega_{GW}\sim 10^{-15}$ with
BBO-lite,  BBO-grand has potential to detect the background close to
$\Omega_{GW}\sim 10^{-18}$ in 10 years.
When we try to measure the amplitude 
and the slope  simultaneously for the spectral shape $\Omega_{GW}(f)=\Omega_{GW,F}(f/F)^n$, their errors can be  highly correlated
and the estimation of the amplitude   might be
degraded, compared with the single parameter fitting for
a flat spectrum $\Omega_{GW}(f)=\Omega_{GW}$.
If we take the central frequency $F$ around the optimal sensitivity for
the correlation analysis, we can nearly  diagonalized the two dimensional
error covariance matrix. 

While we showed  the impacts of the low frequency cut-off associated
with the potential astrophysical foreground that might be difficult to
subtract from the data streams, it is not clear how well we can clean the foreground made by
black hole or neutron star  binaries above $\sim 0.2$Hz. These aspects,
especially in relation to  the correlation analysis,  must be clarified to
properly understand  prospects for the measurement of weak stochastic
background ({\it e.g.} form inflation).
In  addition we have assumed that the detector noises between
combinations $(A,A')$, $(E,E')$ and $(T,T')$  are uncorrelated. In reality they must have
correlations to some degree, and estimation of their magnitude
is  crucial for discussing the sensitivity accessible by  correlation
analysis \cite{Flanagan:1993ix,Allen:1997ad}.

\begin{acknowledgments}
The author thanks S. Kawamura and
S. Phinney  for valuable discussions. He also thanks N. Cornish for
 discussions on astrophysical foreground,  C. Cutler for helpful
 conversations on 
 the noise curves of BBO,   and J. Yokoyama for directing
 his interest to this  research through a collaboration.  

\end{acknowledgments}




\begin{thebibliography}{DUM}


\bibitem{Allen:1996vm}
  B.~Allen,
  arXiv:gr-qc/9604033.



\bibitem{Maggiore:1999vm}
  M.~Maggiore,
  Phys.\ Rept.\  {\bf 331}, 283 (2000)
  [arXiv:gr-qc/9909001].


\bibitem{Seljak:1996gy}
  U.~Seljak and M.~Zaldarriaga,
  Phys.\ Rev.\ Lett.\  {\bf 78}, 2054 (1997)
  [arXiv:astro-ph/9609169].

\bibitem{Kamionkowski:1996zd}
  M.~Kamionkowski, A.~Kosowsky and A.~Stebbins,
  Phys.\ Rev.\ Lett.\  {\bf 78}, 2058 (1997)
  [arXiv:astro-ph/9609132].






\bibitem{Lidsey:1995np}
  J.~E.~Lidsey, A.~R.~Liddle, E.~W.~Kolb, E.~J.~Copeland, T.~Barreiro and M.~Abney,
  Rev.\ Mod.\ Phys.\  {\bf 69}, 373 (1997)
  [arXiv:astro-ph/9508078].





\bibitem{Turner:1996ck}
  M.~S.~Turner,
  Phys.\ Rev.\ D {\bf 55}, 435 (1997)
  [arXiv:astro-ph/9607066].


\bibitem{bbo}
E. S. Phinney et al. The Big Bang Observer, NASA Mission Concept Study (2003).



\bibitem{Seto:2003kc}
  N.~Seto and J.~Yokoyama,
  J.\ Phys.\ Soc.\ Jap.\  {\bf 72}, 3082 (2003)
  [arXiv:gr-qc/0305096].



\bibitem{Ungarelli:2005qb}
  C.~Ungarelli, P.~Corasaniti, R.~A.~Mercer and A.~Vecchio,
  Class.\ Quant.\ Grav.\  {\bf 22}, S955 (2005)
  [arXiv:astro-ph/0504294].


\bibitem{Cooray:2005xr}
  A.~Cooray,
  arXiv:astro-ph/0503118.





\bibitem{Smith:2005mm}
  T.~L.~Smith, M.~Kamionkowski and A.~Cooray,
  arXiv:astro-ph/0506422.

\bibitem{Boyle:2005ug}
  L.~A.~Boyle, P.~J.~Steinhardt and N.~Turok,
  arXiv:astro-ph/0507455;
L.~A.~Boyle and P.~J.~Steinhardt,
  arXiv:astro-ph/0512014.







\bibitem{Farmer:2003pa}
  A.~J.~Farmer and E.~S.~Phinney,
  Mon.\ Not.\ Roy.\ Astron.\ Soc.\  {\bf 346}, 1197 (2003)
  [arXiv:astro-ph/0304393].


\bibitem{Seto:2001qf}
  N.~Seto, S.~Kawamura and T.~Nakamura,
  Phys.\ Rev.\ Lett.\  {\bf 87}, 221103 (2001)
  [arXiv:astro-ph/0108011].


\bibitem{Ungarelli:2000jp}
  C.~Ungarelli and A.~Vecchio,
  Phys.\ Rev.\ D {\bf 63}, 064030 (2001)
  [arXiv:gr-qc/0003021].




\bibitem{Crowder:2005nr}
  J.~Crowder and N.~J.~Cornish,
  arXiv:gr-qc/0506015.



\bibitem{Michelson(1987)}
 P.~F. Michelson,
Mon. Not. R. Astron. Soc.  227, 933 (1997).
 

\bibitem{Christensen:1992wi}
  N.~Christensen,
  Phys.\ Rev.\ D {\bf 46}, 5250 (1992).



\bibitem{Flanagan:1993ix}
  E.~E.~Flanagan,
  Phys.\ Rev.\ D {\bf 48}, 2389 (1993)
  [arXiv:astro-ph/9305029].



\bibitem{Allen:1997ad}
  B.~Allen and J.~D.~Romano,
  Phys.\ Rev.\ D {\bf 59}, 102001 (1999)
  [arXiv:gr-qc/9710117].

\bibitem{estabrook}
F. B. Estabrook and H. D. Wahlquist, Gen Relativ. Gravit {bf 8}, 439
 (1975).

\bibitem{Armstrong et al.(1999)}
J. W.  Armstrong,
F.~B. Estabrook,   and  M. Tinto, Astrophys.J, 527, 814 (1999) 
 



\bibitem{Prince:2002hp}
 T .~A.~Prince, M.~Tinto, .~L.~Larson and .~W.~Armstrong,
  Phys.\ Rev.\ D {\bf 66}, 122002 (2002)
  [arXiv:gr-qc/0209039].




\bibitem{Krolak:2004xp}
  A.~Krolak, M.~Tinto and M.~Vallisneri,
  Phys.\ Rev.\ D {\bf 70}, 022003 (2004)
  [arXiv:gr-qc/0401108].

\bibitem{Nayak:2002ir}
  K.~R.~Nayak, A.~Pai, S.~V.~Dhurandhar and J.~Y.~Vinet,
  Class.\ Quant.\ Grav.\  {\bf 20}, 1217 (2003)
  [arXiv:gr-qc/0210014].

\bibitem{Corbin:2005ny}
  V.~Corbin and N.~J.~Cornish,
  arXiv:gr-qc/0512039.





\bibitem{white}
M. Tinto, F. B. Estabrook, and J. W. Armstrong, LISA Pre-Project
 Publication http://www.srl.caltech.edu/lisa/tdi${}_-$wp/LISA${}_-$Whitepaper.pdf, 2002.

\bibitem{Cornish:2003tz}
  N.~J.~Cornish and R.~W.~Hellings,
  Class.\ Quant.\ Grav.\  {\bf 20}, 4851 (2003)
  [arXiv:gr-qc/0306096].


\bibitem{Shaddock:2003dj}
  D.~A.~Shaddock, M.~Tinto, F.~B.~Estabrook and J.~W.~Armstrong,
  Phys.\ Rev.\ D {\bf 68}, 061303 (2003)
  [arXiv:gr-qc/0307080].

\bibitem{Cutler:1997ta}
  C.~Cutler,
  Phys.\ Rev.\ D {\bf 57}, 7089 (1998)
  [arXiv:gr-qc/9703068].



\bibitem{Tinto:2001ii}
 M.~Tinto, J.~W.~Armstrong and F.~B.~Estabrook,
  Phys.\ Rev.\ D {\bf 63}, 021101 (2001).




\bibitem{Hogan:2001jn}
  C.~J.~Hogan and P.~L.~Bender,
  Phys.\ Rev.\ D {\bf 64}, 062002 (2001)
  [arXiv:astro-ph/0104266].






\bibitem{lisa}
P.~L.~Bender  {\it et al.},
{\it LISA Pre-Phase A Report,} Second edition, July 1998. 




\bibitem{Larson:1999we}
  S.~L.~Larson, W.~A.~Hiscock and R.~W.~Hellings,
  Phys.\ Rev.\ D {\bf 62}, 062001 (2000)
  [arXiv:gr-qc/9909080].


\bibitem{Cornish:2001qi}
  N.~J.~Cornish and S.~L.~Larson,
  Class.\ Quant.\ Grav.\  {\bf 18}, 3473 (2001)
  [arXiv:gr-qc/0103075].






\bibitem{kawamura}
S. Kawamura, private communication (2005).



\bibitem{Seto:2004ji}
  N.~Seto,
  Phys.\ Rev.\ D {\bf 69}, 123005 (2004)
  [arXiv:gr-qc/0403014].





\bibitem{Kudoh:2004he}
  H.~Kudoh and A.~Taruya,
  Phys.\ Rev.\ D {\bf 71}, 024025 (2005)
  [arXiv:gr-qc/0411017];
A.~Taruya and H.~Kudoh,
  Phys.\ Rev.\ D {\bf 72}, 104015 (2005)
  [arXiv:gr-qc/0507114].


\bibitem{Seto:2004np}
  N.~Seto and A.~Cooray,
  Phys.\ Rev.\ D {\bf 70}, 123005 (2004)
  [arXiv:astro-ph/0403259].


\bibitem{Kudoh:2005as}
  H.~Kudoh, A.~Taruya, T.~Hiramatsu and Y.~Himemoto,
  arXiv:gr-qc/0511145.





\bibitem{helstrom}
C. W. Helstrom, Statistical Theory of Signal Detection, 2nd ed. (Pergamon Press, London, 1968).



\bibitem{Cutler:2005qq}
  C.~Cutler and J.~Harms,
  arXiv:gr-qc/0511092.





\end{thebibliography}
\end{document}